\documentclass[useAMS,usenatbib,final]{mn2e}

\usepackage{natbib}
\usepackage{caption}

\bibpunct[]{(}{)}{;}{'a'}{}{,}

\usepackage{graphicx}
\usepackage{amsmath}

\title[A study of Be stars in the Magellanic Clouds]{A study of Be stars in the Magellanic Clouds}
\author[S. Iqbal and S. C. Keller]{S. Iqbal$^{1}$\thanks{E-mail:
shaheen@mso.anu.edu.au} and S. C. Keller$^{1}$\thanks{E-mail: stefan@mso.anu.edu.au}\\
$^{1}$Research School of Astronomy and Astrophysics, The Australian National University, Cotter Road, Weston Creek, ACT 2611, Australia.}
\begin{document}

\date{Received 2013 August}

\pagerange{\pageref{firstpage}--\pageref{lastpage}} \pubyear{2013}

\maketitle

\label{firstpage}

\begin{abstract}
We present the results of a photometric survey for Be stars in eleven young clusters in the Large Magellanic Cloud and fourteen young clusters in the Small Magellanic Cloud. B stars with hydrogen in emission are identified on the basis of their R-H$\alpha$ colour. We find that Be star fraction in clusters decreases with cluster age, and also decreases with the metallicity.
\end{abstract}


\section{Introduction}\label{Sec:intro}

The transient nature of emission lines, especially those of the Balmer series of hydrogen, exhibited in the spectra of some B-type stars is known as the `Be phenomenon'. These spectral changes are attributed to a disk of gaseous material surrounding the central star, the origins of which are still unclear. While recent observational data has allowed for the elimination of some theoretical models \citep[such as the wind compressed disc models of][]{1993ApJ...409..429B}, no satisfactory mechanism for injecting material into the disk has been proposed to date. It is likely that the formation of this disk may be attributed to a number of factors, including, but not limited to, non-radial pulsations, rapid rotation, and the magnetic properties of the central star.

The viscous decretion model suggests that a B star rotating at less than critical velocity ejects material into a circumstellar disk through an unknown mechanism. The material then settles into a Keplarian orbit and the radial structure of the disk is governed by viscosity. It was first proposed by \cite{1991MNRAS.250..432L} and developed by \cite{1997LNP...497..239B}, \cite{2002MNRAS.337..967O} and \cite{2005ASPC..337...75B}.

The Be phenomenon has always been tied to the rapid rotation of a B star. The initial rotational velocity of a B star must be at least 70\% of its critical velocity for it to exhibit Be emission \citep{1996MNRAS.280L..31P,2008A&A...478..467E}. 

To study the possible evolution of B stars into Be stars, rapidly rotating stars have been modelled by \cite{1979ApJ...232..531E} and \cite{1982IAUS...98..299E}. \citeauthor{1979ApJ...232..531E} show that stars that have low rotation at birth are unlikely to develop into Be stars, while those born rotating at 59\% - 76\% of their break up velocity are highly likely to become Be stars. 

However, \cite{2005ApJ...634..585C} suggests that a subset of classical Be stars could could have rotations as low as 40\% - 60\% of their critical velocity. \cite{2011A&A...536A..65D} compare the predictions of evolutionary models for early-type stars for NGC~2004 and the N~11 region on the LMC, and NGC~330 and NGC 346 in the SMC. They find that their nitrogen abundances are inconsistent with those predicted for stars that spend most of the main-sequence lifetimes rotating close to their critical velocity. In particular, they find similar estimated nitrogen enrichment for Be and B type stars, and postulate that either Be stars rotate faster than B stars, but not at critical velocity, or that Be stars only spend a short period (less than 10 \%) of their main-sequence life times rotating close to critical velocity. 

Photometric surveys for Be stars have been conducted in the Galaxy but are limited by the relative paucity of Be stars \citep[e.g.][]{2005ApJS..161..118M,2001AJ....122..248K}. On the other hand, the Magellanic Clouds provide large samples of Be stars in different environments and the opportunity to examine the effect of cluster age and metallicity on the Be phenomenon. Early searches for Be stars in the Magellanic Clouds have been made by \cite{1972MNRAS.159..113F} using objective prisms, while later surveys, such as \cite{1999A&AS..134..489K}, \cite{2006ApJ...652..458W}  and \citet{Martayan2010a}, have relied on photometry achieved through a narrow-band H$\alpha$ filter. Photometric techniques are advantageous as they allow for efficient selection of Be stars, especially within dense clusters where spectroscopy is difficult.  

 \cite{1999A&A...346..459M} summarise the content of Be stars within Large Magellanic Cloud (LMC), Small Magellanic Cloud (SMC) and Milky Way (MW) to disentangle age and metallicity effects of the Be phenomenon. Within their sample there is a clear decrease in Be star fraction with increasing metallicity. Similarly, \cite{2006ApJ...652..458W} use photometry to determine the Be star content of eight clusters in the SMC, five in the LMC, and three in the MW, and find that the Be phenomenon is enhanced in low metallicity environments. Additionally, \cite{2008A&A...478..467E} and \cite{2004PASA...21..310K} find that the fraction of fast rotators on the zero-age main-sequence is higher at lower metallicities. 

This paper examines the effects of metallicity and age on Be star formation through a photometric study of clusters in the Magellanic Clouds. In Section \ref{Sec:obs} we outline the observations undertaken for this study and the data reduction steps used to extract photometry. We describe the selection criteria used to identify Be star candidates in Section \ref{Sec:sel} and compare our results to previous photometric searches for Be stars in Section \ref{Sec:compa}. We outline the results obtained through the photometry in Section \ref{Sec:prop}. Finally we present our conclusions in Section \ref{Sec:conclusion}. 
\vspace{-1.0em}
\section{Observations and data reduction}\label{Sec:obs}
Target clusters were chosen from the OGLE catalog \citep{1996ARA&A..34..419P} satisfying $7 < \log age < 8$, allowing us to probe the Be star content of young clusters. We note that this criterion excludes early-type Be stars from our sample. 

To identify B stars exhibiting hydrogen emission we imaged through a narrow-band H$\alpha$ filter ($\lambda_c =$ 6563 \AA, $\Delta\lambda =$ 300 \AA) and compared it with an image taken in the Cousins $R$ band. Stars with strong hydrogen emission have an $R - H\alpha$ colour that is larger than stars without emission at a given $V - I$. 

The observations were obtained using the Faulkes South Telescope at Siding Spring Observatory from July to November in 2010. Each image is 10.04 $\times$ 10.04 arc minutes in size, with a pixel scale of 0.3 arc seconds per pixel. Exposure times for images using the $g$, $i$ and $R$ filters were 90 seconds each, and three 300 second exposures were taken with the H$\alpha$ filter. While the typical seeing for the entirety of the observing period was between 2" and 3", for each cluster the value varied between 1.83" (ESO~86SC2) and 3.87" (NGC~1902). 

The photometry of the CCD fields was carried out using Source Extractor \citep{1996A&AS..117..393B}. After excluding sources with abnormal Full-Width at Half-Maximum measurements and saturated sources, the stars with good photometry across all four filters were matched. The $g$ and $i$ magnitudes were transformed to Bessell $V$ and $I$ using the Magellanic Clouds Photometric Catalog \cite[MCPC;][]{2004AJ....128.1606Z}, but the R and H$\alpha$ magnitudes were not calibrated and the $R - H\alpha$ colour has an arbitrary zero-point.
\begin{figure}
\centering
\includegraphics[trim=10mm 80mm 10mm 80mm, width=0.50\textwidth]{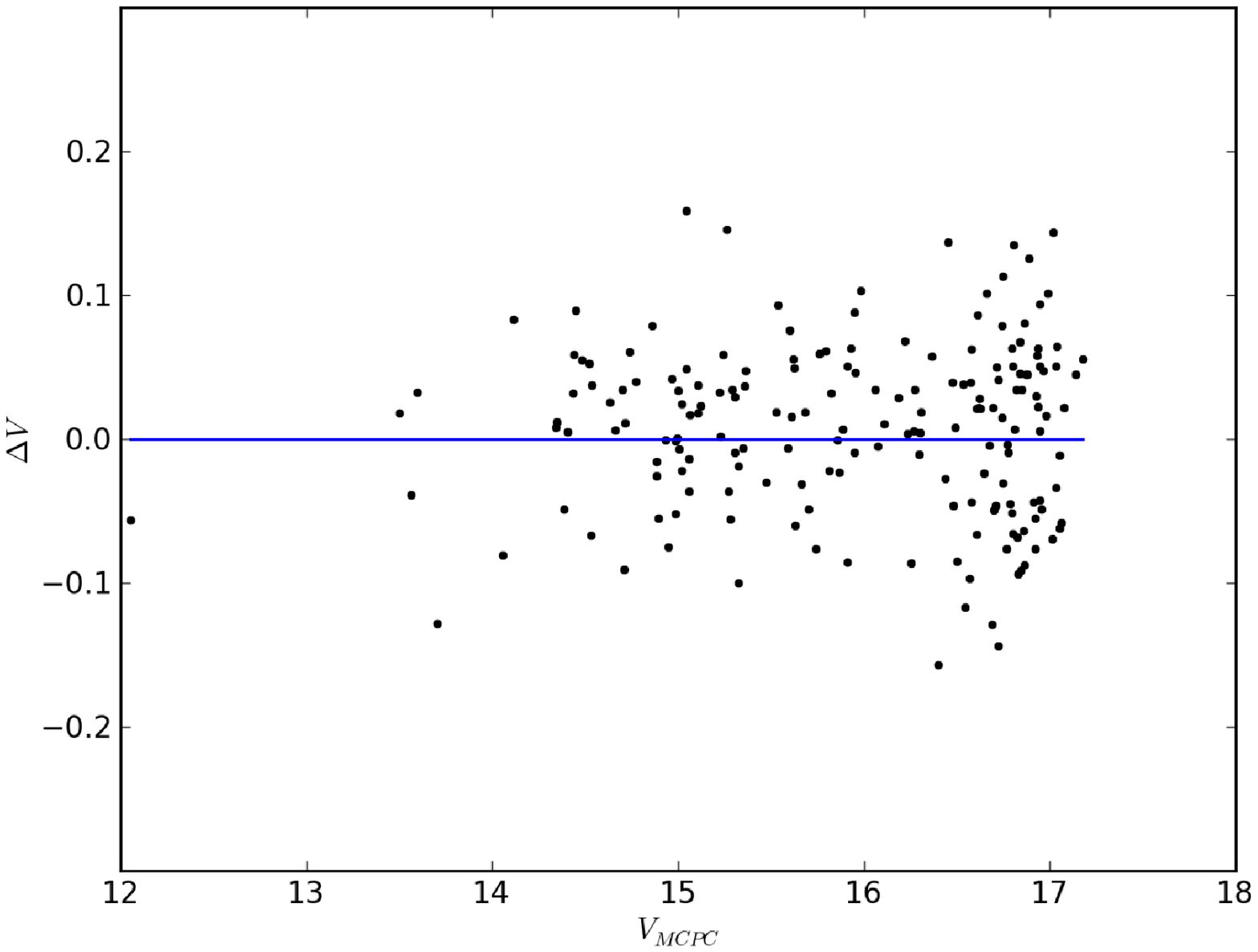}
\caption{The difference in $V$ magnitudes in our work and those in the MCPC for field stars ($I  < 17$) around NGC~330. We noted an offset of 0.02 magnitudes is observed, which we considered in our analysis.}
\label{fig1}
\end{figure}

Figure \ref{fig1} shows the difference in $V$ magnitudes in our work and those in the MCPC for field stars around NGC~330. Only the brightest stars ($I  < 17$) are used to calibrate our data with the MCPC for each cluster. An offset of 0.02 magnitudes is observed, which was taken into consideration in our analysis, while the standard deviation is 0.07 magnitudes. 

In order to determine the cluster members, we find the radius at which the stellar density drops to a value indistinguishable from the field. Stars within this radius are assumed to belong to the cluster, and those outside are treated as belonging to the field population. Typically the innermost region (6 - 12'', typically 10'') of the clusters was excluded from our data due to the difficulty in crossmatching stars across filters, caused by crowding in the \emph{g} filter.
\section{Selection of H$\alpha$ emitting stars}\label{Sec:sel}
\begin{center}
\begin{figure*}
\centering
\includegraphics[trim = 10mm 75mm 10mm 85mm, clip, width=0.8\textwidth]{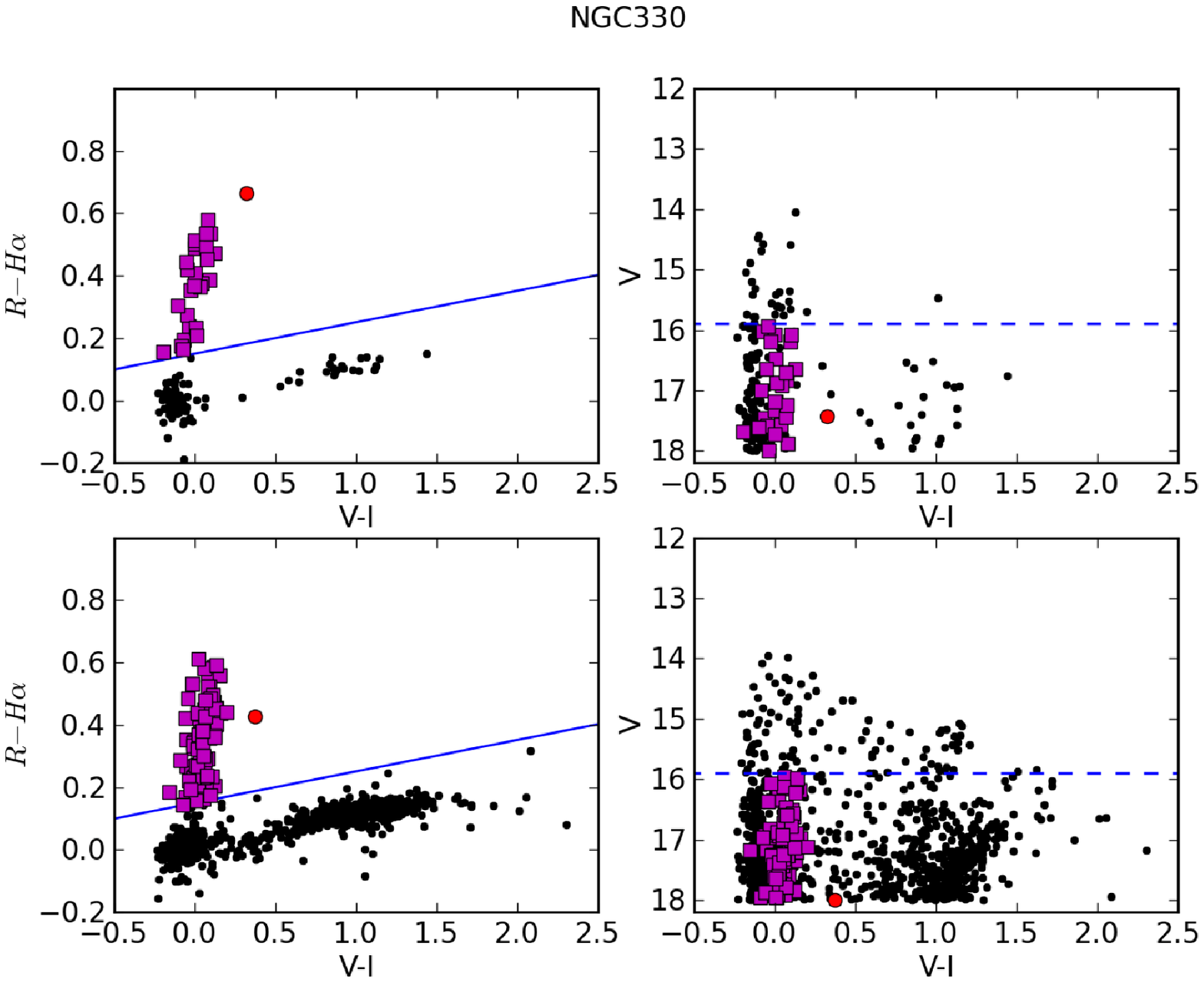}
\caption{The ($R-H\alpha, V-I$) and ($V, V-I$) diagrams for the cluster NGC~330 in the SMC (top) and the surrounding field (bottom). Be stars are identified as the stars above the solid line (blue) in the  ($R-H\alpha, V-I$) diagram and are shown using squares (purple). Note that, as expected for classical Be stars, many candidates lie slightly to the right of the main-sequence. Cooler objects exhibiting hydrogen emission are shown as filled circles (red) and lie above the solid line. The dashed line in the $V, V-I$ diagram indicates the cluster MSTO.}
\label{fig:final_NGC330}
\end{figure*}
\end{center}

\begin{center}
\begin{figure*}
\centering
\includegraphics[trim = 10mm 75mm 10mm 85mm, clip, width=0.8\textwidth]{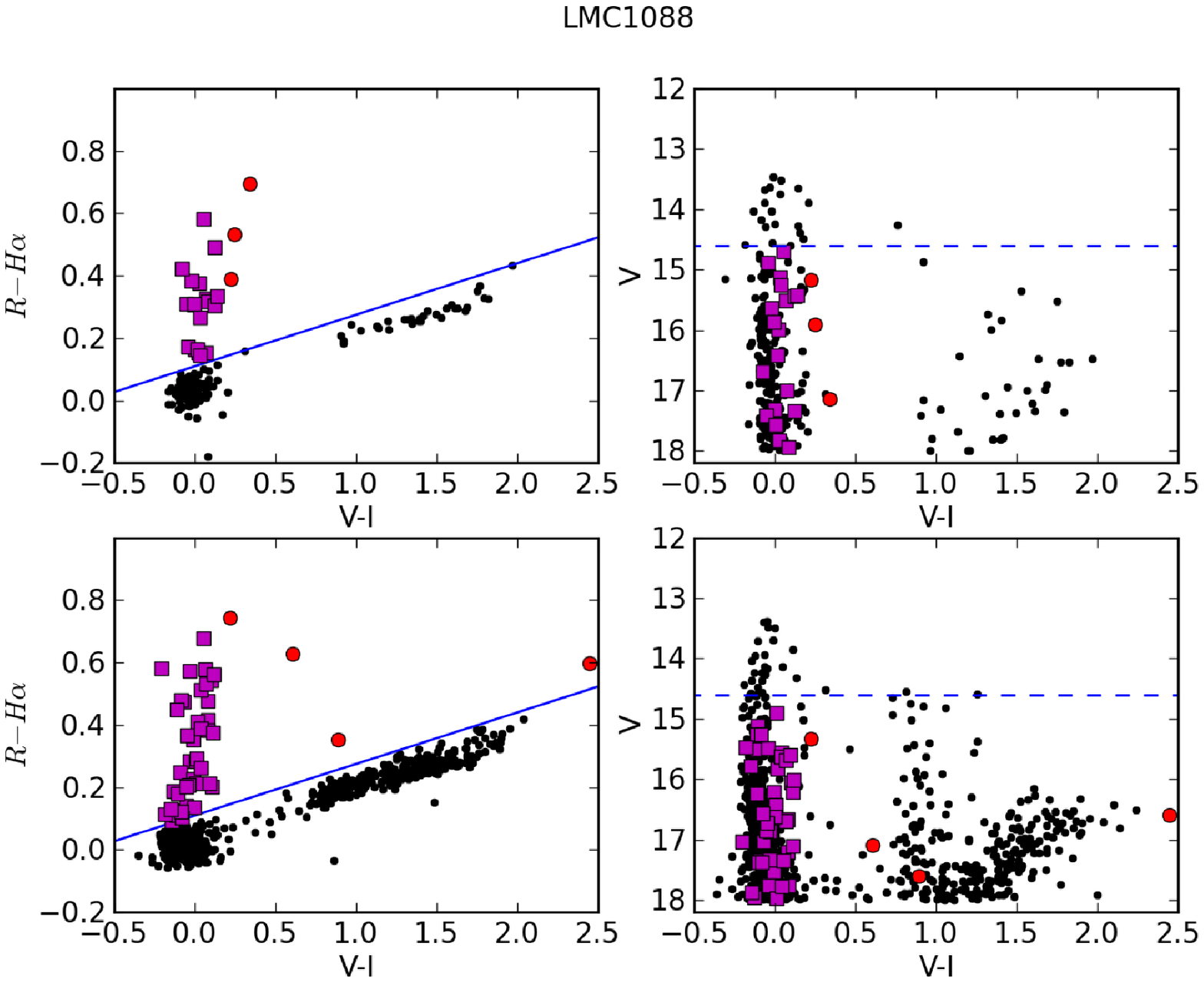}
\caption{Same as Figure \ref{fig:final_NGC330} but for NGC~2004 in the Large Magellanic Cloud}
\label{fig:final_NGC2004}
\end{figure*}
\end{center}

Plots of the colour index $R - H\alpha$ against $V - I$ were made of all stars common to the four filters to identify stars with H$\alpha$ emission. Stars with strong $H\alpha$ emission were selected from these plots. Examples are shown on the left hand side of Figures \ref{fig:final_NGC330} and \ref{fig:final_NGC2004}. 
\begin{center}
\begin{table*}
\captionof{table}{The radii, ages and number of B and Be stars for each cluster and surrounding field in the LMC and SMC}
{\small
\hfill{}
    \begin{tabular}{lcccc | cc | cc | r}
\hline
     & & & & & & & & & \\
     & & & & & Cluster & & Field & & \\
    Cluster &$\log age$ & Radius & Radius & MSTO & $N_{Be}$/$N_B$ & $\underline{N_{Be}}$ & $N_{Be}$/$N_B$ & $\underline{N_{Be}}$ & \textit{N} Field\\
                   &     Myr       &  arc min   & arc min  &  $$V$$ mag &   & $N_B+N_{Be}$& & $N_B+N_{Be}$& Cont. \\
     & & & (from Glatt) & & & & & & \\
\hline
     LMC & & & & & & & & & \\
     & & & & & & & & & \\
NGC 2004	 & 	7.2	 & 	1.85	 & 	1.45	 & 	14.6	 & 	19 (22)/207	 & 	0.084	 & 	42 (43)/827	 & 	0.048 	& 4.64 \\
NGC 1805	 & 	7.6	 & 	0.99	 & 	0.85	 & 	16	 & 	4/46	 & 	0.080	 & 	16/463	 & 	0.033 	& 0.47\\
ESO 86SC2	 & 	7.5	 & 	1.18	 & 	1.1	 & 	15.7	 & 	3/35	 & 	0.079	 & 	12/217	 & 	0.052 	& 0.51\\
NGC 2025	 & 	8	 & 	0.95	 & 	0.95	 & 	17.2	 & 	1/18	 & 	0.053	 & 	21/344	 & 	0.058	& 0.57\\
LMC 0461	 & 	8.1	 & 	0.76	 & 	0.8	 & 	17.5	 & 	0/10	 & 	0.000	 & 	10/113	 & 	0.081 	& 0.17\\
NGC 1755	 & 	7.4	 & 	1.17	 & 	1.02	 & 	15.3	 & 	4/81	 & 	0.047	 & 	17/771	 & 	0.022 	& 0.70 \\
NGC 1774	 & 	7.7	 & 	0.76	 & 	0.8	 & 	16.3	 & 	6/45	 & 	0.118	 & 	10 (11)/447	 & 	0.022	& 0.17\\
NGC 1951	 & 	7.7	 & 	1.25	 & 	0.8	 & 	16.3	 & 	1/32	 & 	0.030	 & 	25/378	 & 	0.062	& 0.19 \\
LMC 0899	 & 	8.1	 & 	1.62	 & 	0.85	 & 	17.5	 & 	0/14	 & 	0.000	 & 	2/89	 & 	0.022 	& 0.17\\
NGC 1902	 & 	8	 & 	1	 & 	0.85	 & 	17.2	 & 	2/10	 & 	0.167	 & 	6/181	 & 	0.032	& 0.18\\
NGC 1839	 & 	7.9	 & 	0.78	 & 	0.8	 & 	16.9	 & 	0/38	 & 	0.000	 & 	39/1220	 & 	0.031 	& 0.70\\
\hline
     SMC & & & & & & & & & \\
     & & & & & & & & & \\
NGC 330	 & 	7.4	 & 	1.18	 & 	1.33		& 	15.9	 & 	27 (28)/136	 & 	0.166 	 & 	80 (81)/976	 & 	0.076 & 3.59\\
NGC 231	 & 	7.9	 & 	0.63	 & 	0.90	 	& 	17.4	 & 	6/14	 & 	0.300 	 & 	47/392	 & 	0.107 & 0.56\\
NGC 220	 & 	8	 & 	1.34	 & 	0.60	 	& 	17.7	 & 	5/11	 & 	0.313 	 & 	21/179	 & 	0.105 & 1.14 \\
SMC 0707& 	7.4	 & 	0.55	 & 	0.42		& 	15.8	 & 	2/16	 & 	0.111 	 & 	146 (148)/1922 & 	0.071 & 1.32\\
SMC 0483& 	7.9	 & 	1.11	 & 	0.60	 	& 	17.4	 & 	0/33	 & 	0.000 	 & 	37 (41)/497	 & 	0.069 & 1.35\\
SMC 0028 & 	8	 & 	1.61	 & 	0.85	 	& 	17.7	 & 	0/11	 & 	0.000 	 & 	4/78	 & 	0.049 & 0.31\\
NGC 306	 & 	7.4	 & 	0.45	 & 	0.55	 	& 	15.8	 & 	1/14	 & 	0.067 	 & 	74/958	 & 	0.072 & 0.44\\
NGC 376	 & 	7.5	 & 	0.69	 & 	0.90	 	& 	16.1	 & 	3/16	 & 	0.158  	 & 	67 (70)/793	 & 	0.078 & 0.97\\
SMC 0720& 	7.6	 & 	0.61	 & 	0.42	 	& 	16.5	 & 	1/9	 & 	0.100 	 & 	42 (43)/737	 & 	0.054 & 0.49\\
SMC 0213& 	7.6	 & 	0.63	 & 	0.60	 	& 	16.5	 & 	1/17	 & 	0.056 	 & 	60 (61)/1119	 & 	0.051 & 0.71\\
SMC 0302& 	7.8	 & 	0.97	 & 	0.43		& 	17.1	 & 	1/26	 & 	0.037 	 & 	38/511	 & 	0.069 & 1.06\\
NGC 299	 & 	7.4	 & 	0.61	 & 	0.45		& 	15.8	 & 	1/9	 & 	0.100 	 & 	55 (57)/757	 & 	0.068 & 0.59\\
SMC 0263& 	7.8	 & 	0.79	 & 	0.48		& 	17.1	 & 	1/13	 & 	0.071 	 & 	39/654	 & 	0.056 & 0.71\\
SMC 0314 & 	7.4	 & 	0.59	 & 	0.55	 	& 	15.8	 & 	0/24	 & 	0.000	 & 	52/795	 & 	0.061 & 0.55\\
\hline
\end{tabular}}
\hfill{}
\textsc{Notes}.--The ages shown are taken from \citet{2010A&A...517A..50G}, and the cluster radii from that work are shown for comparison. The numbers in parentheses include both firm and possible detections.
  \label{tab:res}
\end{table*}
\end{center}
\vspace{-5.0em}
In each cluster we restrict our definition of a B star to stars still on the main-sequence, with $V$ magnitudes between the Main Sequence Turn Off (MSTO) and $V =$ 18. Blue supergiant stars lie directly above the main-sequence stars, with $V < $ MSTO. A star displaying Be characteristics lying above the MSTO is likely to be a supergiant B[e] star: a B type star with forbidden emission lines in its optical spectrum \citep{1998A&A...340..117L}. Since we are only interested in identifying classical (non-supergiant) Be stars in our study, we exclude the stars above the MSTO from our search.

As mentioned in Section \ref{Sec:obs}, the  $R - H\alpha$ colour has an arbitrary zero point. A zero-point for this colour was selected such that the main-sequence stars lie at  $R - H\alpha = 0$ for each cluster. Thus in $R - H\alpha, V - I$ main-sequence stars form a tight clump near the origin, while nearly all the cooler stars form a horizontal band. These stars do not exhibit hydrogen emission. 

The group of stars that extend up and slightly to the red of the main-sequence clump clearly show significant H$\alpha$ emission, and these are the stars of interest to us, shown on the plots as (purple) squares. Occasionally, cooler giant stars also exhibit strong H$\alpha$ emission. Cool objects with excess hydrogen emission are shown as (red) circles on the plot, isolated as potentially interesting objects. Notably, some of these cool objects lie very close to our Be candidates.

The dispersion in $R-H\alpha$ of the normal main-sequence stars in the clump is determined mostly by errors in the H$\alpha$ photometry, which depends on the magnitude of the star. Our Be star selection process is limited to stars with $V < 18$, which corresponds to an uncertainty in H$\alpha$ photometry of 0.25 magnitudes. 

The diagnostic $R-H\alpha, V-I$ diagram for NGC~330 in the SMC is shown in Figure \ref{fig:final_NGC330}, along with the $V, V-I$ colour magnitude diagram (CMD). The same is shown for NGC~2004 in the LMC in Figure \ref{fig:final_NGC2004}. Similar diagrams for each cluster in our sample are included in the online supplementary material. Table \ref{tab:res} summarises the final results of this survey, showing the number of B and Be stars within each cluster and in the surrounding field, and the cluster age as provided by \citet{2010A&A...517A..50G}. Also included are our calculated cluster radii, and for comparison we have included the cluster radii from \citet{2010A&A...517A..50G}. There is generally a good agreement between our values and those of \citeauthor{2010A&A...517A..50G}, who use the apparent major and minor axis adopted by \cite{2008MNRAS.389..678B} for each cluster to compute the mean apparent diameter. 

Additionally, we provide the estimated field contamination of Be stars in each cluster in Table \ref{tab:res}, determined by calculating the number of field Be stars that occur per unit area, and multiplying by the area of the cluster. We note that the number of field contaminants in these clusters is so small that they do not to make a significant impact on the Be star fractions. 
\vspace{-2.0em}
\begin{center}
\begin{figure*}
\centering
\includegraphics[width=0.45\textwidth]{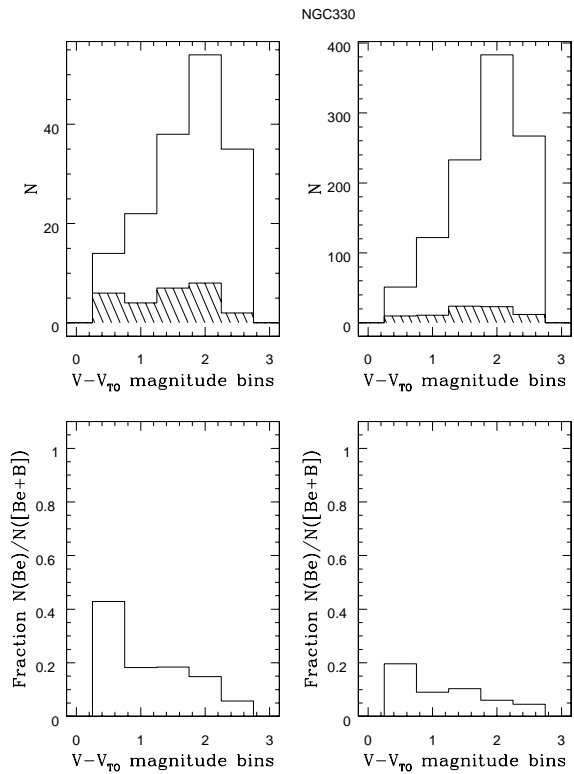}
\caption{Top: Histograms of the number of Be stars (shaded) and main-sequence stars ( $V - I < $ 0.5, including Be stars) in half magnitude bins down the main-sequence of NGC~330 (left) and in the surrounding field (right). Bottom: The ratio of Be stars to main-sequence stars in half magnitude bins down the main-sequence of NGC~330 (left) and in the surrounding field (right).}
\label{fig:hist_NGC330}
\end{figure*}
\end{center}
Using the isochrones of \cite{2000A&AS..141..371G}, we calculate the MSTO for each cluster. We use the distance moduli of the LMC and SMC to arrive at the MSTO points quoted in Table \ref{tab:res}. For the LMC we use a distance modulus of $18.5$ \cite[see][]{2012MNRAS.419.1637L, 2011Ap&SS.tmp..746W,2008MNRAS.390.1762D} and a reddening value of $E(B-V) =$ 0.08. For the SMC we use $18.9$ for the distance modulus \cite[see][]{2009AJ....138.1661S,2010AcA....60..233C} and $E(B-V) =$ 0.05 for reddening. 

The data in Table \ref{tab:res} in parentheses include both firm and possible detections of Be stars. Possible detections are those stars that lie very close to our cutoffs in selecting Be stars, identified by visual inspection of the CMDs and $R-H\alpha, V-I$ diagrams. They are depicted in those diagrams as red circles that lie very close to the main-sequence, and likely belong to the same family of strong emitters. In Figure \ref{fig:final_NGC330}, we identify one possible detection in NGC~330, and one in the surrounding field, while examination of Figure \ref{fig:final_NGC2004} shows three possible detections in NGC~2004, but only one in the field. The other three cool emitters in the field population are rejected as possible detections because they do not lie near the main-sequence.

In identifying possible Be candidates in these clusters, we note that a photometric survey will identify only the stars with large emission equivalent widths, and that a spectroscopic survey with appropriately high resolution will not suffer from this set-back. As such we acknowledge that our results are indicative of the minimum number of Be stars in each cluster; the true Be star fraction can only be higher.

The largest clusters in our sample from the LMC and SMC have been used as templates to investigate the relative Be star frequency between clusters. We use NGC~2004 in the LMC and NGC~330 in the SMC to predict how many Be stars we expect in each cluster. 

For example, consider NGC 220 in the SMC, with a MSTO of 17.7 (see Table \ref{tab:res}). We have 5 Be star candidates in NGC 220, satisfying 17.7 $< V < 18$, which results in a Be star fraction of 0.313. When using NGC~330 as a template, we find that it has 39 stars satisfying 17.7 $< V < 18$, of which 3 are Be stars, resulting in a Be star fraction of 0.076 for the subset. We can thus conclude that NGC 220 is particularly rich in Be stars, when compared to our template cluster, NGC~330. 

This is especially useful in analysing the smallest and the oldest clusters in our sample, which typically contain either a single Be star or none at all. Our comparisons imply that we should not expect these clusters to boast more than one Be star, which is indeed what we find.
\vspace{-1.0em}
\section{Comparison of Be Detections with Previous Works}\label{Sec:compa}
To gauge the robustness of our Be target selections we compare our detections in and around NGC~2004 and NGC~330 to those in \cite{1999A&AS..134..489K}. We restrict our sample to stars with $V < 17$, as they do, in order to make meaningful comparisons.  

We identify 13 Be stars in NGC~2004 with $V < 17$, 9 of which are also found in \cite{1999A&AS..134..489K}. We find four additional candidates within our sample that are not present in their data. They identify 16 Be stars in total, 7 of which we do not detect as Be stars.

Similarly we identify 13 Be stars in NGC~330 with $V < 17$, 11 of which are in common with \cite{1999A&AS..134..489K}. \citeauthor{1999A&AS..134..489K} find a total of 27 Be stars in this cluster, of which 16 are not classified as Be stars in our sample.

We attribute some of the changes in Be star content between the works to the transient nature of the Be phenomenon, but also note that within NGC~330, where the comparison of results seems poorer, many of their candidates occur in crowded regions where our analysis is not as robust due to seeing limitations. 
\vspace{-1.0em}
\section{Results of Photometric Analysis of Be stars}\label{Sec:prop}
\begin{center}
\begin{figure*}
\centering
\includegraphics[width=0.45\textwidth]{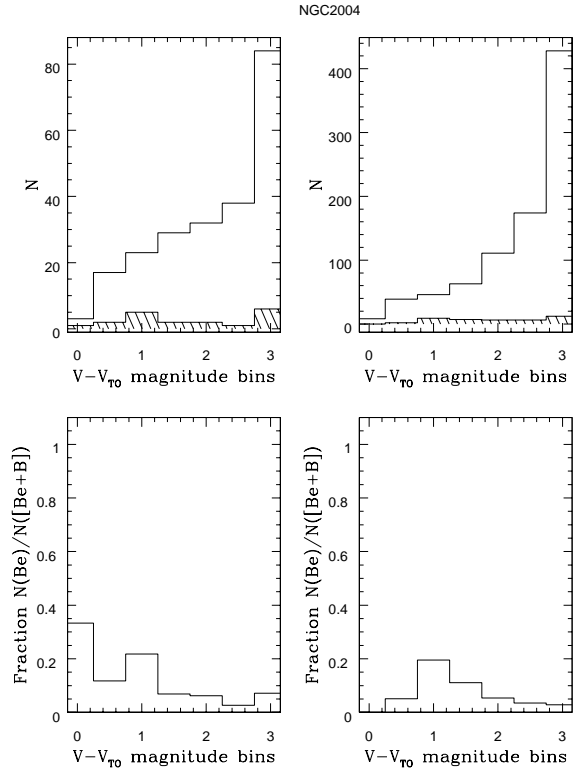}
\caption{Same as Fig. \ref{fig:hist_NGC330} but for NGC~2004}
\label{fig:hist_NGC2004}
\end{figure*}
\end{center}
\vspace{-2.0em}
The Be stars in each cluster appear redder than the general main-sequence population, lying slightly to the right of the main-sequence stars and forming a `red sequence'. This `sequence' can be identified as the Be stars (filled squares) in Figures \ref{fig:final_NGC330} and \ref{fig:final_NGC2004}. We find that the Be stars in the SMC are, on average, redder in $V - I$ by 0.08 than the main-sequence stars, whereas in the LMC the Be stars are redder in $V - I$ by 0.03.

\citet{Martayan2010a} attribute the reddening to rapid rotation and circumstellar disks. Our results are in line with their expectation that reddening effects are stronger in low metallicity environs vs. high metallicity: a consequence of Be stars rotating faster at low metallicity than at high metallicity \citep{2007A&A...462..683M}.  

This has also been investigated by \cite{2012ApJ...756..156H}, who discuss reddening affects due to the scattering of light within the circumstellar disk. In particular, they study where $V$ and $K$ band excesses arise, and find that the disk can be treated as a pseudo-photosphere around the star, with its effective size increasing with wavelength. They note that both the growth and decline rates of $K$ band emission are slower than in the $V$ band, meaning that a star with a circumstellar disk, exhibiting Be characteristics, will appear redder in $V - K$ colour than a main-sequence star. The principle is the same for the $V-I$ colour because of the wavelength dependance of the effect.

\subsection{Be star fraction as a function of luminosity}
Our sample of Be stars allows us to investigate the factors that may influence their creation. A study of the CMDs in Figures \ref{fig:final_NGC330} and \ref{fig:final_NGC2004} shows that Be stars occur over a wide range of luminosities. As they are found at every stage in the main-sequence lifetime, the mechanism producing them is unlikely to be connected to evolutionary phases such as core contraction \citep[see][]{1999A&AS..134..489K}. 

The CMDs in Figures \ref{fig:final_NGC330} and \ref{fig:final_NGC2004} show that the cluster and field populations contain stars of similar age. We investigate the Be star fraction as a function of luminosity by binning both emission line and non-emission line stars within and above the main-sequence clump in the $R - H\alpha, V - I$ plots (corresponding roughly to $V - I < $ 0.5, but varying with each cluster depending on the dispersion in the $V - I $ axis) into 0.5 magnitude bins. The resulting histograms are shown in Figures \ref{fig:hist_NGC330} and \ref{fig:hist_NGC2004} for NGC~330 and NGC~2004 respectively. 

These histograms show the differences in Be star populations between clusters of different ages. We have used ages for each cluster as given by \cite{2010A&A...517A..50G}: $\log age = 7.4$ for NGC~330 and $\log age = 7.2$ for NGC~2004. While the younger cluster, NGC~330, exhibits a higher Be star fraction than NGC~2004 (see Table \ref{tab:res}), the highest fraction of Be stars in each cluster occurs at the MSTO and then declines.

\cite{1999A&AS..134..489K} also observe that Be frequency in young clusters in the Magellanic Clouds reaches a maximum around the main-sequence turn off, whereas the surrounding field has a relatively uniform distribution with luminosity. They argue that this distinction is caused by evolutionary enhancement of rotational velocity occurring late in the main-sequence lifetime. 
\begin{center}
\begin{figure*}
\centering
\includegraphics[width=0.80\textwidth]{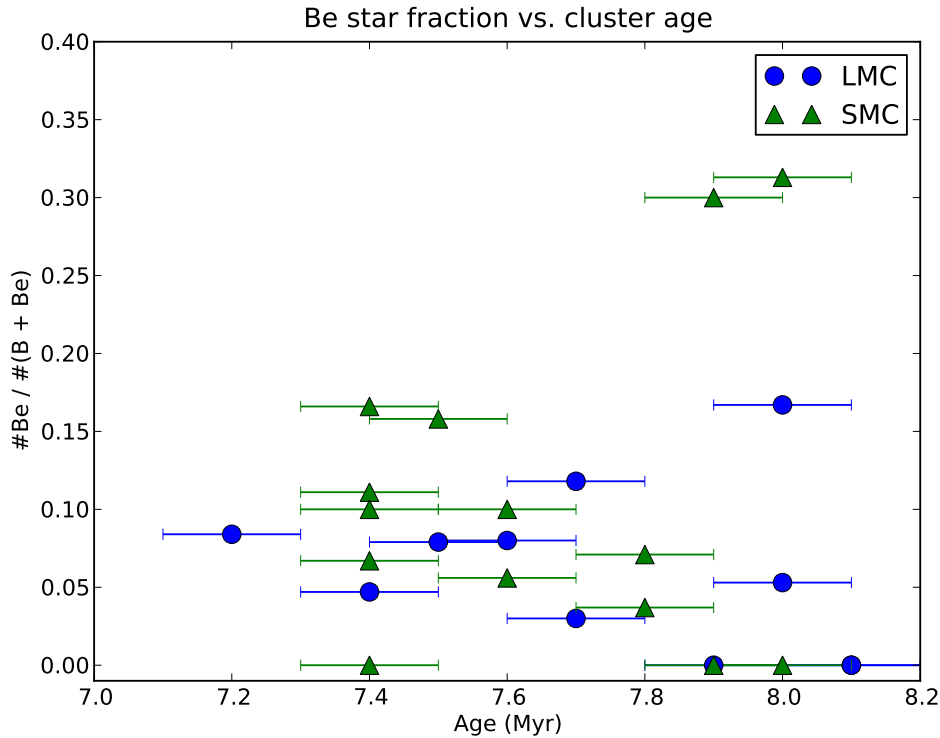}
\caption{The Be star fraction as a function of cluster age for the LMC (blue circles) and SMC (green triangles). The error bars in age correspond to $\log age =$ 0.1 \citep{2010A&A...517A..50G}}
\label{fig:age}
\end{figure*}
\end{center}
\begin{center}
\begin{figure*}
\centering
\includegraphics[width=0.80\textwidth]{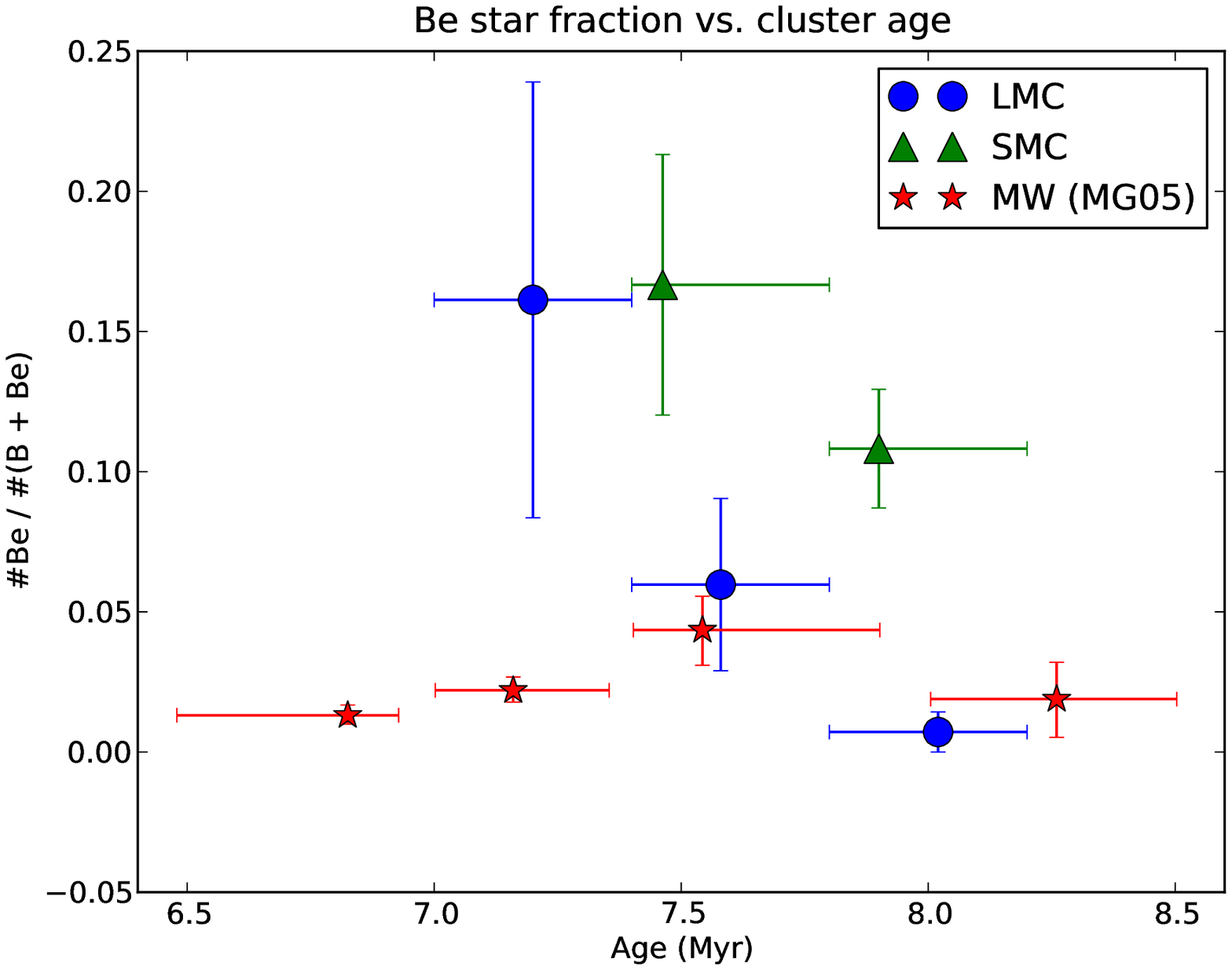}
\caption{The average Be frequency of clusters in the LMC (squares), SMC (triangles) and MW (stars) binned in age:  $7.0 \leq \log age < 7.4$, $7.4 \leq \log age < 7.8$ and $7.8 \leq \log age < 8.2$. The Be frequency is shown with the corresponding Poisson errors, while the error bars in age span the width of the age bins. We only consider stars within one magnitude of the Main Sequence Turn-Off, that is, stars with $V_{MTSO} < V < V_{MSTO} - 1$ for each cluster. The Galactic data is from \citet{2005ApJS..161..118M}. }
\label{fig:age_bins}
\end{figure*}
\end{center}
\vspace{-5.0em}
\subsection{Be star fraction as a function of cluster age}

To study the effect of the age of a cluster on its Be star content we present a plot of Be star fraction as a function of cluster age for our sample in Figure \ref{fig:age}. The Be frequency is shown, and the error bars in age correspond to the $\log age =$ 0.1 that \cite{2010A&A...517A..50G}  claim for their data. No clear relationship is seen in the consideration of the Be star fraction for individual clusters. This is due to the small number statistics in the majority of clusters.  

To better examine the Be star fractions of clusters with different ages, we now consider stars within one magnitude of the MSTO, that is, stars with $V_{MTSO} < V < V_{MSTO} - 1$. 

We plot the average Be frequency in this narrow luminosity range into three age bins for our sample in Figure \ref{fig:age_bins}, as do \cite{2005ApJS..161..118M} with their Galactic sample for easy comparison (also shown in the plot). The bins correspond to ages $7.0 \leq \log age < 7.4$, $7.4 \leq \log age < 7.8$ and $7.8 \leq \log age < 8.2$. Be frequency is shown with the corresponding Poisson errors, while the bars in age now span the width of the age bins. 

The models of \cite{1993ApJ...409..429B} show that the maximum likelihood for the Be phenomenon occurs around spectral type B2, which was confirmed by the results of \citet{1997A&A...318..443Z} in the Galaxy. It is widely held that young stars are the fastest rotators - for example \cite{2008A&A...478..467E} have predicted that stars between 10 and 25 Myrs of age are the fastest rotators, which is also the age range in which most Be stars occur. When convolved with our earlier finding that the Be phenomenon is unconnected to any evolutionary phase, this leads to our finding in Figure \ref{fig:age_bins}, showing a significant decrease in Be star fraction with age in both the LMC and SMC.  

While \cite{2005ApJS..161..118M} claim this trend also appears in their Galactic data, it is more pronounced in the stellar populations of the Magellanic Clouds. They find that the Be star fraction in southern Galactic star clusters increases with age until $\log age = 8$ (100 Myr), with the Be star fraction greatest in clusters within $7.4 \leq \log age < 8.0$, contradicting previous works that find the greatest fraction occurs within  $7.0 \leq \log age < 7.4$. They also find that the Be star fraction decreases with age in LMC clusters older than 100 Myr, a trend that our data also indicates. 

\cite{1982A&A...109...48M} note that Galactic clusters exhibit a trend of decreasing Be star fraction with increasing age, contrasting with the earlier work of \cite{1979ApJ...230..485A} who found no dependance of Be frequency with cluster age. \cite{1999A&AS..134..489K} also find no correlation with age within their sample. 

\citet{Martayan2010a} show that Be stars are more likely to be hosted by younger cluster, with the number of open clusters with Be stars reaching a local minimum towards 30-40 Myr (log (age) = 7.5 - 7.6) and then increasing. Another decrease is noted after 50-60 Myr (log (age) = 7.7 - 7.8). They explain this result by suggesting that some Be stars could be born as be stars, and others have Be characteristics during evolution. They also suggest that the first decrease, which we do not see in our data due to the width of our age bins, could be caused by an evolutionary effect that changes the angular velocity of a born Be star, such that not every star that is born with Be characteristics will remain a Be star throughout its main-sequence lifetime \citep{2007A&A...462..683M}. 
  
 \subsection{Be star fraction as a function of metallicity}
 
Figure \ref{fig:age_bins} also shows a trend of higher Be star fractions with lower metallicity. The highest fractions of Be stars occur in the SMC clusters and decrease as we step with increasing metallicity through the LMC and MW. 

We find that for the stars in the range MSTO $< V < 18$, the Be star fraction in the LMC is 0.06 (Be star fraction varies from 0 to 0.17 ) and in the SMC is 0.11 (varying from 0 to 0.31). Thus, on average, the SMC is 1.8 times richer in Be stars than the LMC. Our results concur with that of \cite{2012MNRAS.421.3622P}, who find that the enhancement of Be stars in the SMC when compared to the LMC varies from 1.4 and 2.6 (depending on the category of Be star as described in their paper). 
 
\cite{1999A&A...346..459M} show that the Be/B fraction in the SMC is about 2.4 times that in our Galaxy and \citet{Martayan2010a} find that, within a comparable age range, early-type Be stars are three to five times more frequent in the SMC ($0.001 < Z < 0.009$) than in the MW ($Z = 0.020$). Citing rapid rotation as the mechanism behind the Be phenomenon, \citet{Martayan2010a} argue that stars in low metallicity environments have higher fractional critical angular rotation rates, $\Omega/\Omega_C$, since they form with the same initial angular momentum as metal-rich stars with the same mass but have smaller radii. Thus the Be star fraction is expected to be higher in the SMC than in the Galaxy. Similarly, \cite{2004PASA...21..310K} finds that Be stars in the LMC cluster and field rotate faster than their Galactic counterparts, indicating a metallicity dependance. 

 \subsection{Rapid rotation as a mechanism behind the Be phenomenon}
 
The number of Be stars peaks at spectral type B2 in both the SMC and Galaxy \citep{Martayan2010a}. The trend we observe with a decreasing number of Be stars as clusters get older can be explained simply by the fact that older clusters no longer host young stars in the spectral range B0 - B2. Interestingly, \cite{2004PASA...21..310K} finds that the fastest rotators in the Galaxy and LMC are found in the spectral range B0 to B2. However, because the maximum strength of the $H\alpha$ emission-line coincides with this spectral type it is questionable whether the Be phenomenon is indeed dependant on effective temperature \citep{Martayan2010a}. The decrease in Be star fraction with increasing metallicity seen in our results may be explained through the increase in rotation seen in stars residing in lower metallicity environs.

 In the context of the Be phenomenon, the first study of the rotational velocities of B and Be stars in the LMC was carried out by \cite{2004PASA...21..310K}, who finds that stars in the LMC rotate faster than those in the MW, providing a possible explanation for the metallicity dependance on the Be star content of a galaxy. \cite{2005ApJS..161..118M} show that up to 73\% of the Be stars in their sample have been spun-up through binary mass transfer, and most of the remaining Be stars were likely rapid rotators since birth. 

\cite{2007A&A...462..683M} investigate the mechanism behind the Be phenomenon further and show that Be stars have higher rotational velocities than B stars at the beginning of their main-sequence lives, and consequently, only a fraction of B stars can become Be stars. \citet{Martayan2010a} claim that the fractional critical angular rotation rate, $\Omega / \Omega_C$ is a fundamental parameter in determining whether a B star develops into a Be star. Importantly, the evolution of the $\Omega / \Omega_C$ parameter is metallicity dependant, but is also different for different mass ranges. Thus, even if rotation is the major cause behind the Be phenomenon, the Be characteristics can manifest themselves under a variety of conditions.  

\cite{2007A&A...462..683M} propose that the Be phenomenon depends on the evolution of the fractional critical angular rotation rate, $\Omega / \Omega_C$. The ratio is governed by evolutionary stage and metallicity, but dependancies vary between mass domains, which causes the discrepancies seen through studies of the Be phenomenon \cite[see][]{2007A&A...462..683M}. The trends we observe with evolutionary stage and metallicity support this theory, however the underlying mechanics of the Be phenomenon remains unclear. 

\section{Conclusion}\label{Sec:conclusion}

We have used a narrow-band H$\alpha$ filter to identify Be star candidates in 11 clusters in the LMC and 14 clusters in the SMC to shed light on the mechanisms behind the Be phenomenon. In particular we focus on the effects of metallicity and age on Be star formation. Our results support the theory, such as that of \citet{Martayan2010a}, that rapid rotation is a key factor in the Be phenomenon.

In particular we note that:
\begin{enumerate}
\item Be stars are redder than B stars in $V - I$ and form a `red sequence'.  \citet{Martayan2010a} predict that reddening effects are stronger in the SMC than in the LMC, and we find this is the case. The Be star sequence in the LMC is 0.03 magnitudes redder than the main-sequence, while in the SMC the separation is 0.08 magnitudes in $V - I$. 
\item The Be star fraction in a cluster peaks at the MSTO luminosity and decreases with decreasing luminosity.
\item Young clusters are more likely to host a larger fraction of Be stars. As cluster age increases the Be star fraction decreases rapidly. This trend is also noted by \cite{2005ApJS..161..118M}. 
\item A low metallicity environment such as the SMC hosts a larger fraction of Be stars that a higher metallicity environment such as the MW, corroborating previous works such as \cite{1999A&A...346..459M},  \cite{2005ApJS..161..118M} and \cite{2004PASA...21..310K}
\end{enumerate}

Young clusters are more likely to host stars of spectral type B0 - B2, which are the fastest rotators in the B class, and thus our result of Be star fraction decreasing with cluster age is a natural consequence. Lower metallicity environments are thought give rise to stars that rotate faster \citep{2007A&A...462..683M} and our result of decreasing Be star fraction with increasing metallicity is in line with this observation.

It is likely that factors aside from rapid rotation, such as magnetism, binarity and pulsations, play major roles in producing Be stars, and that some hither-to unknown mechanism combining these with rapid rotation is the cause of the phenomenon. Our results show that rapid rotation is a major factor in producing Be stars, and affects of the age and metallicity of hosting clusters are closely linked to the rotation of a Be star. It may also be possible that the closer a star comes to critical rotation, the more likely it is to become a Be star, since any of the possible additional mechanisms would not need to be as strong as for a slow rotator.

\section*{Acknowledgments}
This research has made use of NASA's Astrophysics Data System. We thank Prof. Gary Da Costa and our Referee for their insightful comments and suggestions. Shaheen Iqbal acknowledges the support of an Australian Postgraduate Award scholarship. Stefan C. Keller acknowledges the support of the Australian Research Council Discovery Project Grant (DP 120101237).
\bibliographystyle{mn2e}
\bibliography{references}
\bsp

\label{lastpage}

\end{document}